\documentclass[12pt]{article}

\usepackage{epsfig}

\usepackage{color}

\usepackage{graphicx}

\usepackage{amsmath, amsthm}
\usepackage{amscd}
\usepackage{amssymb}
\usepackage{array}
\usepackage{color}
\usepackage{titling}

\newtheorem{thm}{Theorem}[section]
\newtheorem{theorem}[thm]{Theorem}

\theoremstyle{definition}
\newtheorem{definition}[thm]{Definition}
\newtheorem{example}[thm]{Example}

\newtheorem{remark}[thm]{Remark}

\newcommand{\id}{\relax{\rm 1\kern-.28em 1}}

\newcommand{\R}{\mathbb{R}}
\newcommand{\C}{\mathbb{C}}

\newcommand{\beq}{\begin{equation}}
\newcommand{\eeq}{\end{equation}}

\newcommand{\re}{\mathrm{e}}
\newcommand{\ri}{\mathrm{i}}
\newcommand{\rid}{\mathrm{Id}}

\newcommand{\rspan}{\mathrm{span}}
\newcommand{\rSU}{\mathrm{SU}}

\newcommand{\cM}{\mathcal{M}}
\newcommand{\cL}{\mathcal{L}}

\newcommand{\cO}{\mathcal{O}}
\newcommand{\cA}{\mathcal{A}}
\newcommand{\cB}{\mathcal{B}}

\newcommand{\fsu}{\mathfrak{ su}}

\newcommand{\real}{\mathbf{Re}}

\newcommand{\ima}{\mathbf{Im}}
\newcommand{\cH}{\mathcal{H}}
\newcommand{\cS}{\mathcal{S}}
\newcommand{\cU}{\mathcal{U}}

\newcommand{\tr}{\mathrm{tr}}

\begin{document}

\centerline{\Large{\bf
The  fuzzy bit}}

\bigskip
\bigskip

\centerline{ M. Aldana}

\smallskip

\centerline{\it Departamento de Ciencias de la Tierra,
Universidad Sim\'{o}n Bol\'{\i}var}
\centerline {\it
 Valle de Sartenejas, Baruta. Apartado Postal 89000, Venezuela. }

\centerline{{\footnotesize e-mail: maldana@usb.ve}}
\bigskip

\centerline{ M. A. Lled\'{o} }

\smallskip

 \centerline{\it  Departament de F\'{\i}sica Te\`{o}rica,
Universitat de Val\`{e}ncia and}
 \centerline{\it IFIC (CSIC-UVEG)}
 \centerline{\small\it C/Dr.
Moliner, 50, E-46100 Burjassot (Val\`{e}ncia), Spain.}
 \centerline{{\footnotesize e-mail: maria.lledo@ific.uv.es}}

\vskip 1cm

\begin{abstract}

In this paper, the formulation of Quantum Mechanics in terms of fuzzy logic and fuzzy sets is explored. A result by Pykacz, that establishes a correspondence between (quantum) logics (lattices with certain properties) and certain families of fuzzy sets, is applied to the Birkhoff-von Neumann logic, the lattice of projectors of a Hilbert space. Three cases are considered: the qubit, two qubits entangled and a qutrit `nested' inside the two entangled qubits. The membership functions of the fuzzy sets are explicitly computed and  all the connectives of the fuzzy sets are interpreted as operations with these particular  membership functions. In this way, a complete picture of the standard quantum logic in terms of fuzzy sets is obtained for the systems considered.

\end{abstract}

\vfill\eject

\section{Introduction}

In Mackey's approach to Quantum Mechanics \cite{mackey}, the description of a physical system is done in terms of certain axioms. We are given two abstract sets: the set of  {\it states} of the system denoted by $\cS$ and the set of physical quantities or {\it observables}, denoted as  $\cO$. The results of measurements on the physical system are assumed to be real numbers, so one considers $\cB(\R)$,  the family of Borel subsets of the real line. We are given a function
$$\begin{CD}\cO\times \cS\times \cB(\R)@>p>>[0,1]\\
(A,\rho, E)@>>>p(A,\rho, E)\end{CD}$$
that is interpreted as the probability that the measurement of the observable $A$ with the system in the state $\rho$ is contained in the Borel subset $E$. The first six axioms regard  properties of the probability function, $p$.  These axioms  are natural or have physical interpretation. The first conclusion  is that the physical system is described in terms of its {\it logic} $\cL$,  together with a convex family of probability measures on $\cL$. All these terms are going to be defined in Section \ref{logics-sec} (other sources are, for example, \cite{va,bc}).

\bigskip

A particular example of logic is the logic of propositions of a classical system. Propositions are such as `the observable $A$ has a value in $E$'. If $\cS$ is the {\it phase space } or space of states of the classical system, the observables are functions $f:\cS\rightarrow \R$ and the proposition will be true if the system is in a state inside $f^{-1}(E)$. In this way, physically sensible statements about the system are essentially subsets of $\cS$. The logic of Classical Mechanics is the Boolean algebra of subsets of a set $\cS$. This is the standard relation between sets and the propositional logic. Let $U$ be a set and consider the subsets of it. Let $A$ be one such subset and $x$ an element in $U$. Then $A$ can be defined as the collection of elements such that the proposition `$x\in A$' is true.

So the first six axioms apply to Quantum Mechanics, but also to Classical Mechanics

\bigskip

The next axiom of Mackey (Axiom VII) is that the logic of a quantum system is, among all possible logics, the Birkhoff-von Neumann logic of projectors of a Hilbert space.
 Quoting Mackey \cite{mackey}, `This axiom  has a rather different character from Axioms I-VI. These all have some degree of physical naturalness and plausibility. Axiom VII seems completely {\it ad hoc.}'

 \bigskip

In parallel to these logics, the many valued logics of {\L}ukasiewicz \cite{luc} and others, where propositions are not always totally true or false, but have some degree of truth, were being developed. Intuitively, they seem related to a probabilistic theory like Quantum Mechanics.

The same relation that the classical, two valued logic, has with set theory, has the many valued logic with {\it fuzzy sets} \cite{zad}.  A fuzzy subset $\cA$ of a set $\cU$, called the {\it universe of discourse}, is given by a map $\mu_\cA:\cU\rightarrow[0,1]$ called the {\it  membership function}, that associates to any element of $\cU$ a `degree of membership' with respect to $\cA$. The value $\mu_\cA(x)= 1$ means that $x$ belongs totally to $\cA$ and  $\mu_\cA(x)= 0$ means that $x$ does not belong to $\cA$. In between, we can have any degree of membership. If $\cA$ is a `crisp' set, then $\mu_A$ is the characteristic function
$$\mu_\cA=\begin{cases}0\quad \mathrm{if}\, x\notin \cA,\\
1\quad \mathrm{if}\, x\in \cA.
\end{cases}$$
The relationship between fuzzy sets and many valued logic was made explicit by Giles \cite{gi}.

A theorem by Pykacz \cite{py1,py2} makes, finally, the connection between fuzzy sets and Quantum Mechanics. Essentially, it says that giving a logic (in the sense used before) is equivalent to  giving a family of fuzzy sets with certain properties. So any logic does have a `representation' in terms of fuzzy sets. Adding properties to such representation allows to further distinguish between logics. In particular, one can consider the universe of discourse to be the set of density matrices in some Hilbert space, and compute the membership functions in terms of an hermitian operator on such Hilbert space and a Borel subset of $\R$. Then we recover the Birkhoff-von Neumann logic. This choice is, then, equivalent to the choice of Mackey in his Axiom VII.

\bigskip

The aim of the present work is to clarify these abstract concepts, by  giving explicitly the membership functions corresponding to the Birkhoff-von Neumann logic for different systems, all with a finite dimensional Hilbert space. The paper is organized as follows. In Section \ref{logics-sec} we go over the definition of logic, and of  observable and state over that logic. In Section \ref{mac-sec} we deepen the understanding of the axioms of Mackey and the work of M\c{a}czy\'{n}ski \cite{mac} with his {\it experimental functions} (that later will become the membership functions). In Section \ref{QM-sec} we specify all these concepts for the Birkhoff-von Neumann logic. In Section \ref{qubit-sec} we compute the experimental functions for the qubit and in Section \ref{2qubits-sec} we do the same for two qubits entangled, in the formalism of the {\it Bloch matrix} \cite{bk, ga}.
In Section \ref{qutrit-sec} we exploit the decomposition of the Hilbert space of the two entangled qubits in terms of eigenvectors of the angular momentum operator to see how a  qutrit, `nested' in the system of two entangled qubits, arises \cite{bmmsm}  and compute easily its experimental functions in terms of the Bloch matrix. At this point we make a detour to study the unitary transformations of the qutrit in terms of {\it orthogonal symmetric Lie algebras}, inherited from the two entangled qubits \cite{zvsb}. In Section \ref{fuzzy-sec} we turn to the fuzzy set picture. We interpret the  connectives of fuzzy sets for the case of the Birkhoff-von Neumann logic.
Logic gates, essential part of quantum computers, can also be written in terms of the membership functions. We deal with a few examples and an interesting  result is obtained with respect to the square root of NOT. Finally, in Section \ref{conclusion-sec}, we state our conclusions.

\section{Logics}\label{logics-sec}
This section is devoted to give, briefly, the basic definitions of logic, observable and state, following some standard references \cite{va,bc,py2}.

\begin{definition}
	A {\it partially ordered set} or  {\it poset}, is a set $L$ together with a binary relation,   `$\leq$', satisfying
	\begin{itemize}
		\item Reflexivity:  $a \leq a$ for all $a\in L$.
		\item Transitivity: for all $a,b,c \in L$ such that $a\leq b$ and $b \leq c$ then $a \leq c$.
		\item Antisymmetry: if $a, b\in L$ are such that $a\leq b$ and $b\leq a$ then $a=b$.
	\end{itemize}
The binary relation does not need to be defined for all pairs of elements in $L$. When it is so, it is a {\it completely ordered set} or simply an {\it ordered set}.
	
	\hfill$\square$
\end{definition}

\begin{example} \label{powerset3-ex}The set of all the subsets or  {\it power set} of a three element set $\{x, y, z\}$  partially ordered by the inclusion relation is a poset (see Fig. \ref{powerset3}).

\begin{figure}  [htbp]
 \centering
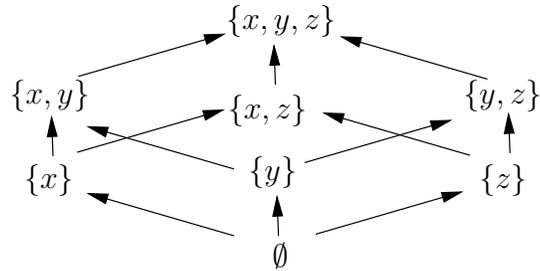
  \caption{Partial order in the power set of a three element set.}
 \label{powerset3}
\end{figure}

\hfill$\square$

\end{example}

\begin{definition}
	If  $L$ is a poset, there exists, at most, one element called {\it zero} and denoted as `0', such that $0 \leq a\,$ $\,\forall a \in L.$ Also, there exists, at most, and element called {\it one} that will be denoted as  `1'  such that $a \leq 1$ $\forall a \in L$.
	
\hfill$\square$
\end{definition}
\noindent In Example \ref{powerset3-ex}, the empty set $\emptyset$ is the 0 and the set $\{x,y,z\}$ is the 1.
 \begin{definition}
 For $S$ a finite subset of  a poset $L$, the {\it meet  of $S$} is, if it exists, an element $\bigwedge S$ that satisfies
\begin{enumerate}
\item $\bigwedge S\leq s$ for all $s\in S$.
\item If $x\in L$ is such that $x\leq s\;$ $\forall s\in S$, then $x\leq \bigwedge S $.
\end{enumerate}
If $S$ has only two elements $a$ and $b$, then the meet will be denoted as $a\wedge b$.

\smallskip

For $S$ and $L$ as above, the {\it join  of $S$} is, if it exists, an element $\bigvee S$ that satisfies:
\begin{enumerate}
\item $s\leq\bigvee S$ for all $s\in S$.
\item If $x \in L$ is such that $s\leq x\;$ $ \forall s\in S$, then  $ \bigvee S\leq x$.
\end{enumerate}
If $S$ has only two elements $a$ and $b$ then the join will be denoted as $a\vee b$.

\hfill$\square$
\end{definition}

The meet and the join, if they exist, are unique.

\begin{definition}A poset $L$ such that
\begin{enumerate}
\item $0$ and $1$ exist, $0\neq 1$,
\item for any finite subset $F\subset L$,  $\bigwedge S$ and $\bigvee S$  exist,
\end{enumerate}
is a {\it lattice}.

\hfill$\square$
\end{definition}

\begin{definition} A {\it complementation} in a lattice  $L$
 is a map
$$\begin{CD}L@>\bot >>L\\
a@>>> a^\bot,\end{CD}$$
such that:
\begin{enumerate}
\item $\bot$ is bijective
\item $a\leq b \Rightarrow b^\bot\leq a^\bot$ (order reversing property)
\item $(a^\bot)^\bot =a$ (involutivity)
\item $a\wedge a^\bot=0$ and $a\vee a^\bot=1$ (complementarity).
\end{enumerate}
\smallskip
If $x\leq y^\bot$, we will say that $x\bot y$.

\hfill$\square$
\end{definition}

\begin{definition}

When a lattice $L$ has a complementation, we say that $L$ is a {\it complemented lattice}.

\hfill$\square$
\end{definition}

\begin{definition}An  {\it orthomodular lattice} is a complemented lattice such that, for any $x, y\in L$ with $x\leq y$ then $y=x\vee(y\wedge x^\bot)$ (orthomodular law).\label{orthomodular}
\hyphenation{ele-ments}

\hyphenation{Axiom}

\end{definition}

\begin{definition} We will say that an orthomodular lattice is {\it $\sigma$-orthocomplete} if, for every  countable set $\{a_l, a_2, \dots , a_n, \dots,\}$ of pairwise orthogonal elements, the join $\bigvee_i a_i$ exists.
\hfill$\square$
\label{sigma-orthocomplete}
\end{definition}

\begin{definition}\label{quantumlogic-def}
We will say that an orthomodular, $\sigma$-orthocomplete lattice is a {\it  logic}.

\hfill$\square$

\end{definition}

\begin{example}{\sl  Boolean algebra of subsets of a set: classical logic.}

Let $L$ be the set of subsets of a certain set $X$ with partial ordering given by inclusion,  join given  by the intersection,  meet by the union and complement by the standard complement in $S$. The meet and the join   satisfy the {\it distributive law}
\begin{align}
	&a\wedge(b\vee c)=(a\wedge b)\vee (a\wedge c),\nonumber\\
	&a\vee(b\wedge c)=(a\vee b)\wedge (a\vee c),\label{distributivity}
	\end{align}
which implies, together with the definition of complement, the orthomodular property. $L$ is a  logic called the {\it classical logic}.

\hfill$\square$

\end{example}

\begin{example}{\sl  Closed subspaces of a Hilbert space: Birkhoff-von Neumann logic \cite{bvn}.}\label{bvn-ex}
\label{BvN}
Let $\cH$ be a Hilbert space,  possibly of infinite dimension, and consider $L $ to be the set of its closed subspaces. There is a partial ordering given by the inclusion; then the meet is the intersection and the join the closure of the sum of vector spaces. The complement is the orthogonal complement inside the Hilbert space.

When the closed subspaces are orthogonal, the join is the direct sum of vector spaces, which is also closed.


One can show that the orthomodular law is  satisfied, thus $L$ is a logic called the {\it standard} or {\it Birkhoff-von Neumann logic}.

$L$ does not satisfy the distributive law, which is the main difference with the classical logic.
\hyphenation{ge-ne-ral}
When the Hilbert space is finite dimensional, all the subspaces are closed and the orthomodular law (Definition \ref{orthomodular}) can be replaced by the more general modular law:
$$x, y, z\in L,\; x\subseteq y,\qquad x\oplus( y\cap z)=(x\oplus z)\cap y.$$

\hfill$\square$

\end{example}

\begin{definition} A probability measure in a logic $L$ is a map $\rho:L\rightarrow [0,1]$  such that
\begin{enumerate}
\item $\rho(1)=1$
\item $\rho(\bigvee_ia_i)= \sum_i \rho(a_i)$ for any sequence of pairwise orthogonal elements of $L$.

\end{enumerate}

\hfill$\square$

\end{definition}
Later, we will interpret a probability measure on the Birkhoff-von Neumann
logic as a state of the system. There is an important property that the states of the system must satisfy:

\begin{definition} A  set of probability measures $\cS$  for a quantum logic $L$ is  said to be {\it ordering} ({\it full, order determining}) if and only if
$$\rho (a)\leq \rho (b)\quad \forall \rho \in \cS\; \Rightarrow\; a\leq b.$$

\hfill$\square$

\end{definition}

\begin{definition} Let $L$ be a logic and $\cB(\R)$ the set of Borel sets of the real line. A map $A:\cB(\R)\rightarrow L$ is an $L$-valued measure if
\begin{enumerate}
\item $A(\emptyset)=0,\qquad A(\R)=1$,
\item   if $E,F\in \cB(\R)$ are such that $E\cap F=\emptyset$ then ,$ A(E)\bot A(F)$,
\item if $E_i\in \cB(\R)$ is a sequence of pairwise  disjoint Borel sets, then $A(E_1\cup E_2\cup \cdots) =A(E_1)\vee A( E_2)\vee \cdots\;\;  $ .

\end{enumerate}

\hfill$\square$

\end{definition}

$L$-valued measures in the Birkhoff-von Neumann
logic play the role of the observables of the quantum theory. Since we can identify a subspace of a Hilbert space with its orthogonal projection, an $L$-valued measure is, in that case, a projection valued measure.

\begin{definition} A  family $M_t, \,t\in T$ of $L$-valued measures is {\it surjective} if for each $f\in L$ there exists $t\in T$ and $E\in\cB(\R)$ such that $M_t(E)=f$.

\hfill$\square$

\end{definition}

\section{The Mackey axioms and the  M\c{a}czy\'{n}ski\\ theorem}\label{mac-sec}

In \cite{mackey}, Mackey describes Quantum Mechanics as a rigorous mathematical system in terms of nine axioms that are, as much as possible, plausible and natural from physical considerations.


One starts with the assumption that, for each quantum system, there exists a set of states $\cS$, a set of observables $\cO$ and a function $p$ such that\begin{equation}\begin{CD}\cO\times\cS\times\cB(\R)@> p>>[0,1]\\
(A, \rho, E)@>>>p(A, \rho, E),\end{CD}\label{probability map}\end{equation}
were $\cB(\R)$ denotes the family of Borel subsets of the real line. The function $p$ has to be interpreted as the probability that the measurement of the observable $A$, when the system is in the state $\rho$,  leads to a value in the Borel set $E$, so
\begin{equation}\begin{CD}\cB(\R)@> \rho_A>>[0,1]\\
E@>>>p(A,\rho,E),
\end{CD}\nonumber\end{equation}
is a probability measure for each pair $\rho, A$. We refer to \cite{mackey} for the explicit formulation of the nine axioms. The first six axioms describe properties of the function $p$, based in physical considerations. We are not detailing those axioms here, as they are stated very clearly in the original reference \cite{mackey}. (Also, for a more modern treatment of the subject, see, for example, \cite{bc}). For example, the $\sigma$-orthocompleteness of Definition \ref{sigma-orthocomplete} is assumed as Axiom V. We  have that $p(A,\rho,E)$ defines a logic in the sense of Definition \ref{quantumlogic-def}.

 Axiom VII is, as we mentioned in the Introduction, {\it ad hoc} and doesn't arise from any natural, physical consideration. It says that the logic of a quantum system is the Birkhoff-von Neumann logic of closed subspaces of a Hilbert space.

Axiom VIII deals with certain observables called {\it questions} (see below) and Axiom IX, with the unitary time evolution of the system.

It is worth to mention here the definition of {\it question} by Mackey. A question is an observable $Q$ such that $\rho_Q$ takes only the values 0 or 1. It corresponds to a proposition that can be answered with `yes' (1) or `no' (0). Let $A$ be any observable and $E$ any Borel set in $\R$. Then the question $Q_A^E$ is the observable that yields 1 whenever a measurement of $A$ is in $E$ and 0 if the measurement of $A$ is not in $E$.

Let $\rho$ be any state and $Q$ any question. Then the function $\rho_Q(\{1\})=s$, $s\in [0,1]$, defines an ordering in the space $L$ of questions: $Q_1\leq Q_2$ if and only if $\rho_{Q_1}(\{1\})\leq \rho_{Q_2}(\{1\})$ for all $\rho\in S$. In particular, it is clear that
\beq \rho_{Q_A^E}(\{1\})=p(A, \rho, E).\label{questions}\eeq

In the space of questions, we have an obvious complementation, $Q^\bot=1-Q$, which takes the value 0 when $Q$ takes the value 1 and vice versa. We will say that two questions are orthogonal if $Q_1\leq Q_2^\bot$, that is,
$$\rho_{Q_1}(\{1\})\leq 1-\rho_{Q_2}(\{1\})\hbox{    or    } \rho_{Q_1}(\{1\})+\rho_{Q_2}(\{1\})\leq 1$$ for each $\rho\in S$.

\smallskip

In the language of M\c{a}czy\'{n}ski \cite{mac}, the functions
\begin{equation}f_A^E(\rho)=p(A,\rho, E)\label{experimentalfunction1}\end{equation} are called {\it experimental functions}. Notice that experimental functions are defined on the equivalence classes of pairs $(A, E)$ under the equivalence relation
$$(A, E)\sim (A', E')\hbox{   if and only if   } p(A,\rho, E)=p(A',\rho, E')\qquad \forall\rho\in S.$$
The experimental function $f_A^E(\rho) $ is associated to the {\it experimental proposition} `The result of the measurement of $A$ lies in $E$', which is one of the questions of Mackey.

%



  In the following, we will consider finite or infinite countable sums of experimental functions, the infinite case meaning that the series is convergent. M\c{a}czy\'{n}ski introduces two convenient definitions:

  \begin{definition}A finite or countable sequence $f_1, f_2, f_3, \dots $ of experimental functions is {\it
  orthogonal} if there exists an experimental function $g$ such that
  $$ f_1+ f_2+ f_3+ \dots  +g=1. $$

  \end{definition}

\begin{definition}  A finite or countable sequence $f_1, f_2, f_3, \dots $ of experimental functions is {\it pairwise orthogonal} if, for each pair $(i,j)$, $i\neq j$, we have that $f_i+f_j\leq 1$.  A one element sequence is, by definition, orthogonal.

\hfill$\square$

\end{definition}

  \hyphenation{or-tho-go-nal}
  Clearly, an orthogonal sequence is pairwise orthogonal, but the contrary is not always true. Nevertheless, if an experimental proposition in an orthogonal sequence is `true' in some state, that is, its experimental function takes the value 1 at such state, all the other experimental functions have to take the value 0, so the rest of the experimental propositions are `false'. This is the same also if the sequence is only pairwise orthogonal. Hence, according to M\c{a}czy\'{n}ski, it makes sense to require that the set of experimental functions satisfies the  postulate that both concepts are equivalent for such set. This is called by  M\c{a}czy\'{n}ski the {\it  orthogonality postulate}. It is equivalent to Mackey's fifth axiom  and it allowed M\c{a}czy\'{n}ski to prove that $L$ is a 
  complemented lattice.
  We have the following theorem by M\c{a}czy\'{n}ski (version adapted by Pykacz \cite{py2}):

  \begin{theorem} (M\c{a}czy\'{n}ski'73 \cite{mac})\label{mac-theorem}
  Let $\cO$, $\cS$ and $p$ be given as above, let $L$ be the set of experimental functions defined by $p$ and assume that they satisfy the orthogonality postulate. Then,
\hyphenation{na-tu-ral}
  \begin{enumerate}

  \item $L$ is a logic with respect to the natural order of real functions, with complementation
  $f^\bot=1-f$, $\,f\in L$.

  \item Each observable $A$ determines a unique $L$-valued measure $M_A:\cB(\R)\rightarrow L$ defined as
  $$M_A(E)=f_A^E\,,\qquad \forall E\in \cB(\R).$$
  The family $\{M_A, A\in \cO\}$ is surjective

  \item Each state $\rho\in \cS$ determines a unique probability measure $m_\rho:L\rightarrow [0,1]$ defined as
  $$m_\rho (f)= f(\rho)\qquad \forall f\in L.$$
  The family  $\{m_\rho, \rho\in \cS\}$ is ordering.

  \item For each $A\in \cO$, $\rho \in \cS$ and $E\in \cB(\R)$
  $$p(A, \rho, E)=m_\rho\circ M_A(E).$$
  \end{enumerate}

  Conversely, if $L$ is a logic  admitting an ordering set of probability measures $\cS$ and $\cO$ is a surjective set of $L$-valued measures, then the function defined by
  $$p(A, \rho, E):=\rho\circ A(E)$$
  with $\rho \in\cS$ and $A\in \cO$ is a probability function that satisfies the orthogonality postulate and whose logic, $L^f$, of $p$ is isomorphic\footnote{Two logics are isomorphic if there is a bijection  between them preserving the order relation and the orthocomplementation.} to $L$.

  \hfill$\square$

  \end{theorem}

\section{Quantum Mechanics}\label{QM-sec}
The VIIth axiom of Mackey says that the quantum logic is isomorphic to the Birkhoff-von Neumann logic $\cL$ of closed subspaces (or the projectors over such subspaces) of a certain separable Hilbert space $\cH$. We saw in Example \ref{BvN} that the partial ordering is given by the inclusion, the meet is the intersection and the join is the sum. The complement is the orthogonal complement inside the Hilbert space with inner product $\langle\,\cdot,\,\cdot\,\rangle$.

We follow closely the argument of Mackey \cite{mackey}. Let $\psi$ be a unit vector on the Hilbert space and let $P$ be a projector. It is clear that
$$\begin{CD}P@>>>m_\psi(P)=\langle P\psi,\psi\rangle\end{CD}$$
is a probability measure on $\cL$ representing the pure state $\psi$. One can prove that a convex linear combination of these probability measures
$$m=\sum_{i=1}^Na_im_{\psi_i}, \qquad a_i\in \R^+,\qquad \sum_{i=1}^Na_i=1\,$$  defines also a probability measure on $\cL$, generically corresponding to a mixed state. Now, by a theorem of Gleason \cite{gl}, one can also prove that in dimension bigger than 2, these are all the probability measures if we assume the reasonable condition that, for each non zero projector, there exists at least a vector $\phi\in \cH$ such that $m_\phi(P)\neq 0$.
\hyphenation{ne-ga-ti-ve}
Observables are defined as $\cL$ valued measures, that is, projection valued measures. By the spectral theorem,  there is a one to one correspondence between projection valued measures and self-adjoint operators. Hence observables correspond to self-adjoint operators.

\hyphenation{ope-ra-tor}
We will focuss on finite dimensional Hilbert spaces. General (mixed or pure) states are described by {\it density matrices}, that is, self-adjoint, non negative matrices with trace 1. If $\rho$ is a density matrix, $A$ a self-adjoint operator and $E$ a Borel subset of $\R$, the probability function $p(A, E, \rho)$  is the trace of the corresponding projector times the density matrix. In any case, $0\leq p(A, E, \rho)\leq 1$, since in the standard formulation of Quantum Mechanics this quantity is a probability. We will extract some consequences from this fact.

\section{The qubit}\label{qubit-sec}

We want to describe the qubit system with the BvN (Hilbert space  or standard) logic in terms of M\c{a}czy\'{n}ski functional representation of Quantum Mechanics.

We then consider the Hilbert space $\cH=\C^2$. The set of observables is the set of self-adjoint operators
$$\Theta=\left \{\, A\in M_{2\times 2}(\C)\; |\; A=A^\dag\,\right\}.$$ In the basis of the Pauli matrices
$$\sigma_0=\rid=\begin{pmatrix}1&0\\0&1\end{pmatrix},\quad \sigma_1=\begin{pmatrix}0&1\\1&0\end{pmatrix}, \quad \sigma_2=\begin{pmatrix}0&-\ri\\\ri&0\end{pmatrix},\quad \sigma_3=\begin{pmatrix}1&0\\0&-1\end{pmatrix}$$ a self-adjoint operator $A$ can be written as
$A=\sum_{\mu=0}^3a_\mu\sigma_\mu\;$, $a_\mu\in \R$
\beq A=\begin{pmatrix}a_0+a_3&a_1-\ri a_2\\a_1+\ri a_2&a_0-a_3\end{pmatrix},\qquad \det A=a_0^2-a_1^2-a_2^2-a_3^2.\label{observable}\eeq

\begin{figure}  [htbp]
 \centering
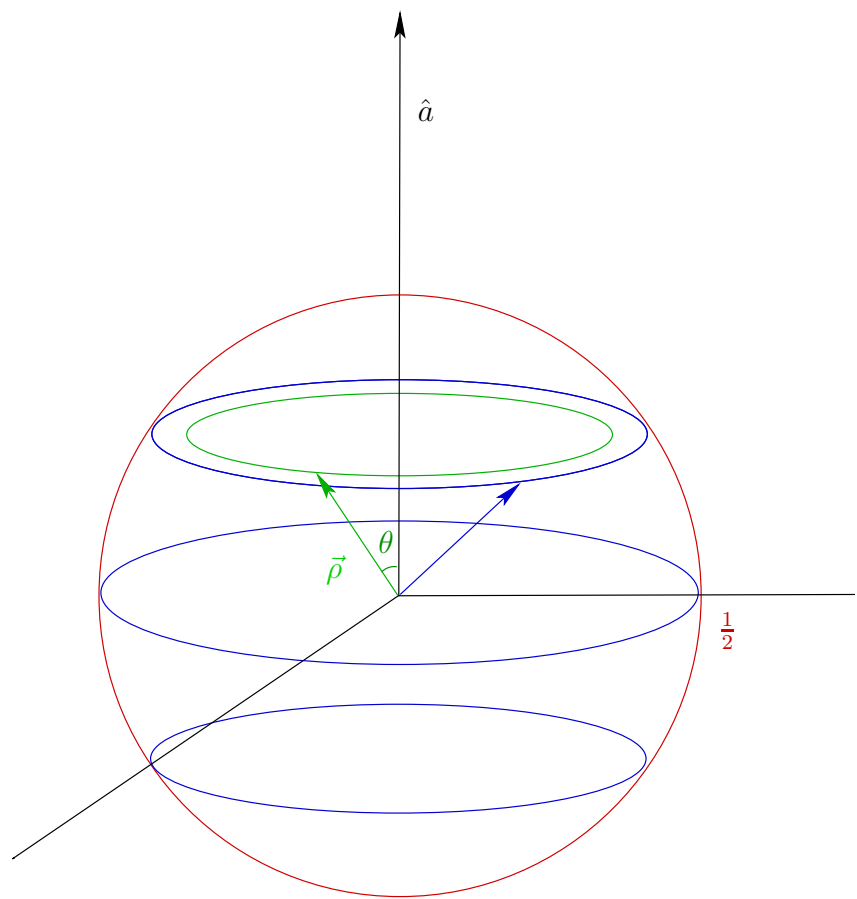
  \caption{Bloch ball of states and vector associated to the observable \cite{MATHEMATICA}.}
 \label{Bloch-fig}
\end{figure}

The set of states $\cS$ consists of the density matrices, that is,  $2\times 2$  self-adjoint matrices that are non negative (the eigenvalues are non negative) and with trace 1:
$$\cS=\left \{\, \rho\in M_{2\times 2}(\C)\; |\; \rho=\rho^\dag,\, \rho\hbox{ is non negative and } \tr\rho=1\,\right\}.$$
We will write a density matrix as
\begin{equation}\rho = \begin{pmatrix}\rho_0+\rho_3&\rho_1-\ri \rho_2\\\rho_1+\ri \rho_2&\rho_0-\rho_3\end{pmatrix}=\rho_0\rid+\vec{\rho}\cdot\vec{\sigma}\\,\label{densitymatrix}\end{equation} with $\rho_\mu$ real.
The trace condition implies that $\rho_0=1/2$. From the characteristic equation
$$\det (\rho-\lambda \rid)=0$$ we obtain
$$ \rho_\pm=\rho_0\pm\sqrt{\rho_1^2+\rho_2^2+\rho_3^2},\label{eigenvalues}$$
and taking into account that $\rho_{\pm}>0$, the non negativity of $\rho$ is equivalent to  $\det\rho\geq0$, that is,
 \beq\rho_1^2+\rho_2^2+\rho_3^2\leq \frac 14 ,\label{density}\eeq so they can be represented as a ball of radius 1/2, the {\it Bloch ball} (see Fig.\ref{Bloch-fig})

 The pure states correspond to density matrices satisfying
 $$\tr\rho^2=1.$$ They
  form  a sphere of radius 1/2, the {\it Bloch sphere}. For these states the equality in (\ref{density}) holds. Antipodal points represent mutually orthogonal states. The points inside the Bloch sphere represent mixed states and its  center represents the maximally mixed state.

The eigenvalues of $A$ in (\ref{observable}) are also
\beq\lambda_\pm =a_0\pm |\vec{a} |, \qquad \vec{a}=(a_1,a_2,a_3),\label{eigenvaluesA}\eeq  and the eigenvectors   are given by  the conditions
\begin{align}
&\vec{v}=\begin{pmatrix}v_1\\v_2\end{pmatrix}\nonumber\\
&\lambda_+:\qquad (a_3-|\vec{a}|)v_1=-(a_1-\ri a_2)v_2\nonumber\\
&\lambda_-:\qquad (a_1+\ri a_2)v_1=(a_3-|\vec{a}|) v_2\label{eigenvectorsA}
\end{align}

Let $\cB(\R)$ denote the set of Borel subsets of the real line. We then have four types of Borel sets with respect to the eigenvalues $\lambda_\pm$:
\begin{itemize}
\item Borel sets $E_0$ such that $\lambda_\pm \notin E_0$,

\item Borel sets $E_+$ such that $\lambda_+\in E_+$ and $\lambda_-\notin E_+$,

\item Borel sets $E_-$ such that $\lambda_-\in E_-$ and $\lambda_+\notin E_-$,

\item Borel sets $E_\pm$ such that $\lambda_+,\lambda_-\in E_\pm$.

\end{itemize}
The projector $P_A^E$ on the subspace of eigenvectors in $\cH$ that correspond to eigenvalues contained in $E$  is the same inside of each of the above four classes. We have:
\begin{align}&P^0_A=0, \qquad &&P^+_A=\frac {1}{2|\vec{a}|}\begin{pmatrix}|\vec{a}|+a_3&a_1-\ri a_2\\a_1+\ri a_2&|\vec{a}|-a_3\end{pmatrix}, \nonumber\\\nonumber\\
&P^\pm_A=\rid,\qquad &&P^-_A=\frac {1}{2|\vec{a}|}\begin{pmatrix}|\vec{a}|-a_3&-a_1+\ri a_2\\-a_1-\ri a_2&|\vec{a}|+a_3\end{pmatrix} .\label{projectorsA}\end{align}

\begin{remark} If the two eigenvalues coincide, then $|\vec{a}|=0$ so $a_1=a_2=a_3=0$, the observable is just $a_0\rid$ and the projector is the identity. In that case only $E_0$ and $E_\pm$ are possible.
\hfill$\square$\label{remarkeignevalues}\end{remark}

The experimental functions:
$$\begin{CD}\cS @>f_A^E>>[0,1]\\
\rho@>>>f_A^E(\rho)=p(A, E, \rho),\end{CD}$$
represent  the probability that the outcome of the measurement of the observable $A$ in a state $\rho$ is in the Borel subset $E$. This is computed as the trace of the projector times the density matrix $\tr(P_A^E\rho)$:

\begin{align*}&f^0_A(\rho)=0, \qquad &&f^+_A(\rho)=\frac {1}{2} +\frac{1}{|\vec{a}|}(a_1\rho_1+a_2\rho_2+a_3\rho_3), \\\\
&f^\pm_A(\rho)=1,\qquad &&f^-_A(\rho)=\frac {1}{2} -\frac{1}{|\vec{a}|}(a_1\rho_1+a_2\rho_2+a_3\rho_3)\,\end{align*}
Notice that changing $\vec{a}$ by $-\vec{a}$ we switch from $f^+_A$ to $f^-_A$, so if $A$ is arbitrary it will be enough to consider the functions, say,
$f^+_A$.

Indeed, we see that the experimental functions, other than 0 and 1, depend only on the unit vector $\hat{a}=\vec{a}/|\vec{a}|$, with $\vec{a}=(a_1, a_2,a_3)$. We will denote those functions as
\begin{equation} f_{\hat{a}}(\vec{\rho})=\frac 12 +\hat{a}\cdot \vec{\rho}=\frac 12 +|\vec{\rho}|\cos\theta\label{experimentalfunction},\end{equation} where
$\vec{\rho}=(\rho_1, \rho_2,\rho_3)$, $|\vec{\rho}|\leq 1/2$ (see (\ref{density})) and $\theta$ is the angle between $\vec{\rho}$ and $\hat{a}$. There is no dependence on the azimutal angle.

Let  us consider the vector $\hat{a}$ fixed. The function $f_{\hat{a}}$  only depends on  $\cos \theta$ and $|\vec{\rho}|$. When $|\vec{\rho}|=1/2$ we have a pure state and  the set of pure states with $\theta$ constant is represented in the Bloch sphere of Fig. \ref{Bloch-fig} by the blue circles. The green vector represents an arbitrary state. One can check directly that $0\leq f_{\hat{a}}(\vec{\rho})\leq 1$.

\begin{figure}  [htbp]
 \centering
 \includegraphics[width=100mm]{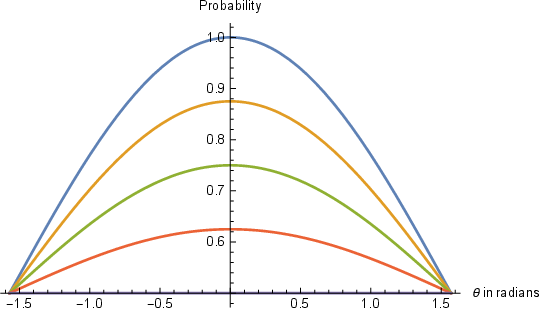}
 \caption{Experimental functions of the qubit in terms of the angle between the Bloch vector and the unitary vector associated to the observable $A$ \cite{MATHEMATICA}.}
 \label{Experimental functions-fig}
\end{figure}

If we fix $|\vec{\rho}|$,  we  obtain the functions $f_{\hat{a}}$ in terms of $\theta$ only. In Fig. \ref{Experimental functions-fig} we have represented some of these functions for different values of $|\vec{\rho}|$. The function for  pure states $|\vec{\rho}|=1/2$ corresponds to the blue curve.

The orthogonality condition is
 \begin{align}f_{\hat{a}}+f_{\hat{a}'} =1+(\hat{a}+\hat{a}')\cdot \vec{\rho}\leq 1\quad \Rightarrow\quad (\hat{a}+\hat{a}')\cdot \vec{\rho}\leq  0,\label{orthogonalityqubit}\end{align}
 which is only possible, for all $\vec{\rho}$, if $\hat{a}=-\hat{a}'$.

  We thus conclude that the only types of pairwise orthogonal sequences are:
\begin{enumerate}
\item $\{ 0, 1 \}$
\item$\{0, f_{\hat{a}}\}$
\item $\{ f_{\hat{a}}, f_{-\hat{a}}\}$
\item $\{0, f_{\hat{a}}, f_{-\hat{a}}\}$
\end{enumerate}
We recall here that a one element sequence is, by definition, orthogonal and that (\ref{density}) implies that $f_{\hat{a}}(\rho)\leq 1$  (as it should be for a probability). We observe then, that all pairwise orthogonal sequences are also orthogonal sequences, so the orthogonality postulate is satisfied.

We can restrict the experimental functions to pure states only. Let
$$\left\{|0\rangle,|1\rangle\right\}$$ be the standard basis in $\cH$, and let $\Psi\rangle \in \cH$. We choose
$$|\Psi\rangle=\Psi_0|0\rangle+\Psi_1|1\rangle\in \cH\qquad \hbox{with}\qquad \langle \Psi|\Psi\rangle=|\Psi_0|^2+|\Psi_1|^2=1 \,.$$  Then
$$f_{\hat a}^+(\Psi_0,\Psi_1)=\frac 12+\frac 1{|\vec{a}|}\left(a_1\real(\Psi_0\bar\Psi_1)-a_2\ima(\Psi_0\bar\Psi_1)+a_3\frac {|\Psi_0|^2-|\Psi_1|^2}{2}\right)\,.$$ If $\Psi_0$ and $\Psi_1$ are real, then we can parametrize the state  as
$$\Psi_0=\cos\alpha,\qquad \Psi_1=\sin\alpha,\qquad \alpha\in [0, 2\pi[$$ and the experimental functions take the simple form
$$f_{\hat a}(\alpha)=\frac 12+\frac 1{|\vec{a}|}\left( a_1\cos\alpha\sin\alpha+a_3\frac 12(\cos^2\alpha-\sin^2\alpha)\right)\,. $$ If we choose  a reference frame in which only $a_3=1$ and the rest 0, we get
$$f_{\hat a}(\alpha)=\cos^2\alpha\,.$$
(This equation holds even if there is a phase between $\Psi_0$ and $\Psi_1$).

\section {Two qubits entangled.} \label{2qubits-sec}
We have now a Hilbert space that is a tensor product of the Hilbert spaces of  the two qubits, $\cH\simeq\C^2\otimes \C^2$.  We will work  with the basis of the matrices $\cM_{2\times 2}(\C)$ formed as products of the Pauli matrices (Dirac basis):
\beq\{\sigma_\mu\otimes \sigma_\nu\}_{\mu,\nu=0}^3\label{diracmatrices}\eeq
We will use Einstein's convention, so sum over repeated indices should be understood. Greek letters will represent indices that run from 0 to 3, while Latin letters will represent indices that run from 1 to 3.  We will use the notation for  the density matrix in terms of the
{\it Bloch matrix} $[r_{\mu\nu}]$ (see \cite{bk})
\beq \rho=\frac 14 r_{\mu\nu}\sigma_\mu\otimes\sigma_\nu,\qquad
[r_{\mu\nu}]=\begin{pmatrix}1&\vec{r}^{\,t}\\\vec{s}& R\end{pmatrix},\label{decomposition} \eeq
which is real. One has immediately that $\tr \rho=1 $ since $\tr \sigma_i=0$ . The vectors $\vec{s}$ and $\vec{r}$ can be interpreted as the  Bloch vectors of the first  and second (respectively) qubits. In fact, tracing out (see property 3 below) the second (first) subsystem, one obtains
\beq\rho_1=\tr_2\rho=\frac 12\rid+ \vec{s}\cdot\vec{\sigma},\qquad \rho_2=\tr_1\rho=\frac 12\rid+ \vec{r}\cdot\vec{\sigma},\label{traceout}\eeq which comparing with (\ref{densitymatrix}) gives
\beq\vec{s}=\vec{\rho_1},\qquad \vec{r}=\vec{\rho_2}.\label{traceout2}\eeq
 $R=[r_{ij}]$ is called the {\it correlation matrix}.

 Using the fact that
$\tr(\sigma_\mu\sigma_\nu)=2\delta _{\mu\nu}$ and the property 3 below, one has that
$$r_{\mu\nu}=\tr(\rho\cdot\sigma_{\mu}\otimes \sigma_\nu).$$

The eigenvalues of the sum of two operators have no easy expression in terms of the eigenvalues of each operator, so we will restrict our analysis to factorizable operators, that is, operators of the form $C=A\otimes B$. We recall that:
\begin{enumerate}
\item $C^\dagger =A^\dagger\otimes B^\dagger$.
\item $\det(A\otimes B)=(\det A)^2(\det B)^2$.
\item $\tr (A\otimes B)=\tr A \cdot\tr B$
\item If $\lambda$ is an eigenvalue of $A$ with eigenvectors $u^a$, where $a$ runs over the multiplicity of $\lambda$, and $\mu$ is an eigenvalue of $B$ with eigenvectors $v^b$, $b$ running on the multiplicity of $\mu$, then $\lambda\mu$ is an eigenvalue of $A\otimes B$ and $u^a\otimes v^b$ are eigenvectors of $\lambda\mu$.
\end{enumerate}
According to (\ref{eigenvectorsA}), we have  eigenvalues $\lambda_\pm=a_0\pm|\vec{a}|$ for $A$ and $\mu_\pm=b_0\pm|\vec{b}|$ for $B$, so we have 4  eigenvalues (some of them may coincide) for $A\otimes B$, $$\alpha_{++}=\lambda_+\mu_+,\quad\alpha_{+-}=\lambda_+\mu_-,\quad\alpha_{-+}
=\lambda_-\mu_+,\quad\alpha_{--}=\lambda_-\mu_-.$$

\begin{example}
Let us consider the simple case $$A=B=\sigma_3=\begin{pmatrix}1&0\\0&-1\end{pmatrix}.$$
Then
$$
C=\sigma_3\otimes\sigma_3=\begin{pmatrix}
1&0&0&0\\
0&-1&0&0\\
0&0&-1&0\\
0&0&0&1
\end{pmatrix},
$$
which has two eigenvalues, $\pm1$, each one with degeneracy 2. Given the operator $C$, the decomposition $C=A\otimes B$ is not unique, since, by the properties of the tensor product, one can rescale $A'=r A$, $B'=B/r$ and $C=A'\otimes B'$ too. It may happen that there is a natural choice, as for this example. Otherwise we have to choose one such decomposition. The operators $A\otimes \rid$ and $\rid \otimes B$ determine, in this case, uniquely a unitary vector of the standard basis by giving their simultaneous  eigenvalues, that is, $\lambda=\pm 1$ of $A$, $\mu=\pm1$ of $ B$. Whenever there is no possibility of confusion, we will denote (as usual)
$$A\otimes \rid \rightarrow A,\qquad \rid \otimes B\rightarrow B.$$

Operators $A$ and $B$ represent local properties of the system, that is, properties of each qubit.

\hfill$\square$
\end{example}

So we will assume that we have chosen  one such decomposition $C=A\otimes B$, since we are limiting our study to decomposable observables. The following analysis will have an equivalent one for every such decomposition.

\medskip

It is useful to classify the Borel sets according to the eigenvalues $\lambda_\pm,\mu_\pm$ that they enclose.
There are the following types of Borel sets (for each type the projector is similar):
\begin{enumerate}
\item Sets that do not contain any eigenvalue.
\item Sets that contain only one eigenvalue.
\item Sets that contain  only the two eigenvalues of the same subsystem.
\item Sets that contain only  two eigenvalues, one of each subsystem.
\item Sets that contain only three eigenvalues.
\item Sets containing the four eigenvalues.
\end{enumerate}
(See Remark \ref{remarkeignevalues} for the case in which one of the operators has a degenerate eigenvalue. This would reduce the number of possibilities.)

We compute the experimental functions for each type of Borel set:

\begin{enumerate}

\item For type 1 we just have $P_0\otimes P_0=0$.

\item Type 2 corresponds to a  projector $P_0$  in one subsystem times a projector over the eigenspace of  $\lambda_\pm$ ($\mu_{\pm}$) in the other. There are four posible combinations but the result is always 0.

\item Type 3 corresponds to $\rid\otimes P_0=0,P_0\otimes\rid=0.$

\item For type 4, let us consider $P_{A\otimes B}^{+,+}=P^+_A\otimes P^+_B$, the projector onto the subspace of eigenvectors in $\cH$ that correspond to the eigenvalues $\lambda_+,\mu_+$, that is, we are considering a Borel subset $E$ that contains only $\lambda_+$ and $\mu_+$ where, according to (\ref{projectorsA}), we have:
$$P^+_A=\frac {1}{2|\vec{a}|}\begin{pmatrix}|\vec{a}|+a_3&a_1-\ri a_2\\a_1+\ri a_2&|\vec{a}|-a_3\end{pmatrix},\quad P^+_B=\frac {1}{2|\vec{b}|}\begin{pmatrix}|\vec{b}|+b_3&b_1-\ri b_2\\b_1+\ri b_2&|\vec{b}|-ba_3\end{pmatrix}.$$ We denote as  $\hat{a}$, $\hat{b}$ the respective unit vectors for $\vec{a}$ and $\vec{b}$. The experimental functions on an arbitrary density matrix (\ref{decomposition}) can be explicitly computed as:
$$f^{+,+}_{\hat{a},\hat{b}}(\rho)=\tr (\rho P_{A\otimes B}^{+,+}).$$  After some algebra we get
\beq f^{+,+}_{\hat{a},\hat{b}}(\vec{s},\vec{r}, R)=\frac 14 \left( 1+\vec{s}\cdot \hat{a}+\vec{r}\cdot \hat{b}+\hat{a}^{\, t}R\hat{b}\right).\label{f++}\eeq
This can be read as the probability for the observables $A$ and $B$ getting the values $\lambda_+$ and $\mu_+$, respectively. The observable $A\otimes B$ gets the value $\alpha_{++}=\lambda_+\mu_+$.

Unlike for the qubit, it is not  straightforward to check here that $0\leq f^{+,+}_{\hat{a},\hat{b}}(\vec{s},\vec{r}, R)\leq 1$, but since this is a probability, according to the Hilbert space machinery, the inequalities must be satisfied. At the end of the section we will check it for pure states. These inequalities should be related to the positivity conditions for the matrix $\rho$, obtained, in the formalism of the Bloch matrix, in \cite{ga}.

To get the rest of the experimental functions for type 3, we only have to change in the projectors  $\hat{a}$ for $-\hat{a}$ to obtain $\lambda_-$ and  $\hat{b}$ for $-\hat{b}$ to obtain $\mu_-$.

\item Type 5 are projectors of the form $P^{+-,\pm}=\rid\otimes P^\pm_B$ or $P^{\pm,+-}=P^\pm_A\otimes\rid$. The  experimental function of, say, $P^{+,+-}=P^+_A\otimes\rid$ reproduces the result of (\ref{experimentalfunction})
\beq f^{+,+-}_{\hat{a}}(\vec{s})=\frac 12 +\vec{s}\cdot\hat{a}.\label{subsystem}\eeq

\item Type 6 is just the identity.
\end{enumerate}

A series of experimental functions is orthogonal if its sum is less or equal than one. This happens, for all density matrices, when we sum probabilities of events without intersection. For example,
 \begin{equation}f^{+,+}_{\hat{a},\hat b}(\vec{s}, \vec{r}, R)+ f^{+,-}_{\hat{a},\hat{b}}(\vec{s},\vec{r}, R)=\frac 12 +\vec{s}\cdot\hat{a}\leq 1,
\label{inequality00} \end{equation}
 according to the Hilbert space formalism. We can extract, as a consequence, the inequalities
 $$-\frac 12\leq\vec{s}\cdot\hat{a}\leq \frac12.$$ In particular, choosing appropriately  $\hat{a}$, we get
 $$-\frac 12\leq s_i\leq \frac12.$$
  The same can be said about the  components of  $\vec{r}$.

Another case is the following sum
\beq f^{+,+}_{\hat{a},\hat{b}}(\vec{s},\vec{r},R)+f^{-,-}_{\hat{a},\hat{b}}(\vec{s},\vec{r},R)=
  \frac 12+\frac 12 \left({\hat{a}}^t R \hat{b}\right)\leq 1,\label{inequality0}\eeq  which implies
 $$-1\leq {\hat{a}}^t R \hat{b}\leq 1.$$ In particular, choosing the axes $\hat{a}$ and $\hat{b}$ appropriately, we have that all the entries of the correlation matrix satisfy
 $$-1\leq R_{ij}\leq 1.$$
 We can also choose $\hat{a}=(1, 0, 0)$ and leave $\hat{b}$ arbitrary; then:
$${\hat{a}}^t R \hat{b}=(R_{11},R_{12}, R_{13})\cdot\begin{pmatrix} b_1\\b_2\\b_3\end{pmatrix}=\left(R_{11}^2+R_{12}^2+ R_{13}^2\right)^{1/2}\cos \beta\leq 1,$$ where $\beta$ is the angle between the vectors $(R_{11},R_{12}, R_{13})$ and $\hat{b}$. This implies
$$R_{11}^2+R_{12}^2+ R_{13}^2\leq 1.$$
The same reasoning can be done with $\hat{a}=(0,1,0)$ or  $\hat{a}=(0,0,1)$. We then obtain that the vectors corresponding to the rows of $R$, $(R_{i1}, R_{i2}, R_{i3})$, satisfy inequalities
\beq R_{i1}^2+R_{i2}^2+ R_{i3}^2\leq 1,\qquad i=1,2,3.\label{inequality1}\eeq
If instead we leave $\hat{a}$ arbitrary and choose for $\hat{b}$ the unit vectors of the three perpendicular axes, we will obtain that the columns of $R$ satisfy  similar inequalities,
\beq R_{1i}^2+R_{2i}^2+ R_{3i}^2\leq 1\qquad i=1,2,3.\label{inequality2}\eeq

Again, (\ref{inequality1}) and (\ref{inequality2})  are inequalities constraining the space of the density matrices, independently of any observable.

In \cite{ga} the conditions of positivity for the matrix $\rho$ are investigated in terms of the decomposition that we  are using here. One such condition is
$$\tr(R^tR)+|\vec{s}|^2+|\vec{r}|^2\leq 3.$$ But

\begin{align*}\tr(R^tR)=&&
\left(R_{11}^2+R_{21}^2+R_{31}^2\right)+
\left(R_{12}^2+R_{22}^2+R_{32}^2\right)+
\left(R_{13}^2+R_{23}^2+R_{33}^2\right)=\\
&&\left(R_{11}^2+R_{12}^2+R_{13}^2\right)+
\left(R_{21}^2+R_{22}^2+R_{23}^2\right)+
\left(R_{31}^2+R_{32}^2+R_{33}^2\right),\end{align*}hence, conditions (\ref{inequality1}) and (\ref{inequality2}) imply, in particular, that $\tr(R^tR)\leq 3$, which is consistent
with the fact that $|\vec{s}|^2+|\vec{r}|^2\geq 0$.
For the case where the equalities in (\ref{inequality1}) and (\ref{inequality2}) are fulfilled, one has that $\tr(R^tR)=3$, which implies that $|\vec{s}|=|\vec{r}|=0$ so
$\vec{s}=\vec{r}=0$. This condition is known as the two qubits being {\it locally maximally mixed}, i.e.,  when tracing out one of the subsystems, the other is left in a maximally mixed state, so its  density matrix is proportional to  the identity (see Equation (\ref{traceout})).

\paragraph{Pure states.}Let $\{|u_1\rangle,|u_2\rangle\}$  and $\{|v_1\rangle,|v_2\rangle\}$ be standard basis in $\C^2$  and let $ |\psi\rangle\in \cH=\C^2\otimes \C^2$ be a unit vector representing a pure state:

\beq|\psi\rangle=\sum_{i,j=1,2}\lambda_{ij}|u_i\rangle\otimes|v_j\rangle,\qquad
\langle\psi|\psi\rangle=\sum_{i,j=1,2}|\lambda_{ij}|^2=1,\label{purestate}\eeq
for $\lambda_{ij}\in \C$. As before, we consider the observable $A\otimes B$ with
$$A=\begin{pmatrix}a_0+a_3&a_1-\ri a_2\\a_1+\ri a_2&a_0-a_3\end{pmatrix},\qquad
B=\begin{pmatrix}b_0+b_3&b_1-\ri b_2\\b_1+\ri b_2&b_0-b_3\end{pmatrix}.$$ We also have the  projectors onto the eigenspaces of $\lambda_+$,$\,\mu_+$ (see (\ref{eigenvaluesA})),  given in terms of the unitary vectors
\begin{align*}&\hat{a}=\frac 1a(a_1, a_2,a_3), \qquad&& a^2=a_1^2+ a_2^2+a_3^2,\\
&\hat{b}=\frac 1b(b_1, b_2,b_3), \qquad&& b^2=b_1^2+ b_2^2+b_3^2,\end{align*}

$$P_A^+=\frac12\begin{pmatrix}1+\hat{a}_3&\hat{a}_1-\ri\hat{a}_2\\
\hat{a}_1+\ri\hat{a}_2&1- \hat{a}_3\end{pmatrix}, \qquad P_B^+=\frac12\begin{pmatrix}1+\hat{b}_3&\hat{b}_1-\ri\hat{b}_2\\
\hat{b}_1+\ri\hat{b}_2&1- \hat{b}_3\end{pmatrix}.$$
We consider now the case $\hat a=(0,0,1)$. Notice that, by a change of the reference frame,  we can always choose our basis in the three dimensional space, such that this condition is  satisfied. Therefore  we are not loosing generality. $B$ is still general. Then we have
$$A=\begin{pmatrix}a_0+1&0\\0&a_0-1\end{pmatrix},\qquad P_A^+=\begin{pmatrix}1&0\\0&0\end{pmatrix}. $$

We compute the probability that a measurement of an observable $A\otimes B$ gives the eigenvalue $\lambda_+\mu_+$. To facilitate the writing, we make some definitions:

\begin{align*}
&\lambda_1:=\lambda_{11},\qquad \lambda_2:=\lambda_{12}\\
&\vec{\lambda}:=\left(\lambda_1 \lambda^*_2+\lambda_2\lambda^*_1,
-\ri(\lambda_1\lambda^*_2-\lambda_2\lambda^*_1),|\lambda_1|^2-|\lambda_2|^2\right).
\end{align*}

One can prove that $|\vec{\lambda}|=\sqrt{\lambda_1|^2+|\lambda_2|^2}$. Then it results that

\begin{align*}&\tilde{f}_{\hat a,\hat b}^{+,+}=\langle\psi|P_A^+\otimes P_B^+|\psi\rangle=
\frac 1{2}\left(|\lambda_1|^2+|\lambda_2|^2+\hat b\cdot\vec{\lambda}\right)
\end{align*}
Let $\theta$ be the angle between $\vec{\lambda}$ and $\hat{b}$. Since, according to (\ref{purestate}), $|\lambda_1|^2+|\lambda_2|^2\leq 1$,

$$\tilde{f}_{\hat a,\hat b}^{+,+}\leq\frac 12 (|\lambda_1|^2+|\lambda_2|^2)^{1/2}(1+\cos \theta)\leq 1\,. $$

 The function $\tilde{f}_{\hat a,\hat b}^{+,+}$ is the restriction to pure states of the function ${f}_{\hat a,\hat b}^{+,+}$ in (\ref{f++}) and we are able to check directly that it satisfies the inequalities
 $$0\leq \tilde{f}_{\hat a,\hat b}^{+,+}\leq 1.$$

\hfill$\square$

\section{The `nested' qutrit.} \label{qutrit-sec} In $\C^2$ we consider the spin operators
$$S_i=\frac12 \sigma_i,\qquad \qquad (\hbar=1).$$
In $\cH=\C^2\otimes \C^2$ we consider the spin operators $S_{1z}=S_z\otimes \rid$ and $S_{2z}=\rid\otimes S_z$.  The {\it standard basis} is the ordered basis of eigenvectors of these operators and it is denoted as
\begin{align}&|00\rangle=|0\rangle\otimes |0\rangle,\quad |01\rangle=|0\rangle\otimes |1\rangle, \nonumber\\ &|10\rangle=|1\rangle\otimes |0\rangle,\quad |11\rangle=|1\rangle\otimes |1\rangle,\label{standardbasis}\end{align} where the indices $0,1$ denote the eigenvalues $+ 1/2$ and $-1/2$, respectively.

The total angular momentum $\vec{S}=\vec{S}_1+\vec{S}_2$, modulus squared, is given in terms of this basis as
$$\vec{S}\cdot \vec{S}=S_{x}^2+ S_{y}^2+ S_{z}^2=
\begin{pmatrix}1&0&0&0\\
0&1/2&1/2&0\\
0&1/2&1/2&0\\
0&0&0&1
\end{pmatrix},$$
with eigenvalues $\lambda=0$ (multiplicity 1) and  $\lambda=1$ (multiplicity 3). A basis of eigenvectors is
\begin{align}&|00\rangle,\quad |q_s\rangle=\frac 1{\sqrt{2}}\big(|01\rangle+|10\rangle\big),\quad |11\rangle,\qquad (\mathrm{eigenvalue} \, 1),\nonumber\\&\mathrm{and}\nonumber
\\
&|q_a\rangle=\frac 1{\sqrt{2}}\big(|01\rangle-|10\rangle\big)\qquad (\mathrm{eigenvalue} \,0).
\label{qutritbasis}\end{align}
Following \cite{bmmsm}, we denote this basis as the {\it entangled basis}.
\hyphenation{fac-to-ri-zed}

As in (\ref{decomposition}), we express the density matrix of the entangled qubits in the standard basis as

$$\rho^{\mathrm{stn}}=\begin{pmatrix}
\rho_{11}&\rho_{12}&\rho_{13}&\rho_{44}\\
\rho_{12}^*&\rho_{22}&\rho_{23}&\rho_{24}\\
\rho_{13}^*&\rho_{23}^*&\rho_{33}&\rho_{34}\\
\rho_{14}^*&\rho_{24}^*&\rho_{34}^*&\rho_{44}\\
\end{pmatrix}=\frac14r_{\mu\nu}\sigma_\mu\otimes \sigma_\nu,\qquad [r_{\mu\nu}]=\begin{pmatrix}1&\vec{r}^{\,t}\\\vec{s}& R\end{pmatrix}$$
Developing the last term, containing the Bloch matrix, one finds (see \cite{ga})
$${ \rho^{\mathrm{stn}}=
\frac14}
\begin{pmatrix}
{\scriptstyle 1 +R_{33}+s_3+r_3}&{\scriptstyle R_{31}-\ri R_{32}+r_1-\ri r_2}&
{\scriptstyle R_{13}-\ri R_{23}+s_1-\ri s_2} & {\scriptstyle R_{11}-\ri R_{12}-\ri R_{21}-R_{22}}\\
{\scriptstyle R_{31}+\ri R_{32}+r_1+\ri r_2}&{\scriptstyle 1-R_{33}+s_3-r_3} & {\scriptstyle R_{11}+\ri R_{12}-\ri R_{21}+R_{22}}&{\scriptstyle -R_{13}+\ri R_{23}+s_1-\ri s_2}\\
{\scriptstyle R_{13}+\ri R_{23} +s_1+\ri s_2}&{\scriptstyle R_{11}-\ri R_{12}+\ri R_{21}+R_{22}}&
{\scriptstyle 1-R_{33}-s_3+r_3}&
{\scriptstyle-R_{31}+\ri R_{32}+r_1-\ri r_2}\\
{\scriptstyle R_{11}+\ri R_{12}+\ri R_{21}-R_{22}}&{\scriptstyle -R_{13}-\ri R_{23}+s_1+\ri s_2}
&{\scriptstyle -R_{31}-\ri R_{32}+r_1+\ri r_2}&{\scriptstyle 1+R_{33}-s_3-r_3}
\end{pmatrix}.
$$

To change the basis from the standard to the entangled one, we  compose the vertical array of unit vectors with the unitary matrix $A$:
$$\begin{pmatrix}
|00\rangle\\
|q_s\rangle\\
|11\rangle\\|q_a\rangle
\end{pmatrix}=A\begin{pmatrix}
|00\rangle\\
|01\rangle\\
|10\rangle\\|11\rangle
\end{pmatrix},\quad A=\begin{pmatrix}1&0&0&0\\
0&1/\sqrt{2}&1/\sqrt{2}&0\\
0&0&0&1\\
0&1/\sqrt{2}&-1/\sqrt{2}&0\end{pmatrix}.$$
Then, the density matrix in the entangled basis is $\rho^\mathrm{ntgl}=A^\dagger\rho^{\mathrm{stn}}A$. We will denote it as
$$\rho^\mathrm{ntgl}=\begin{pmatrix}\tau_{11}&\tau_{12}&\tau_{13}&\beta\\
\tau_{12}^*&\tau_{22}&\tau_{23}&\gamma\\
\tau_{13}^*&\tau_{23}^*&\tau_{33}&\delta\\
\beta^*&\gamma^*&\delta^*&\alpha
\end{pmatrix},\qquad \tau_{11}+\tau_{22}+\tau_{33}+\alpha=1. $$
One could prepare a state with $\alpha=\delta=\gamma=\beta=0$. Then, it represents a  qutrit inside the two entangled qubits. In \cite{bmmsm}, this `nested' qutrit is studied in the context of the dipolar coupling of the two qubits.  There, a family of Hamiltonians are given, for which  time evolution of the system preserves the constraints defining the qutrit.

Writing the constraints in terms of the Bloch matrix, one can see that the conditions for the qutrit become simply

\beq\vec{r}=\vec{s},\qquad R=R^t.\label{qutritcondition}\eeq
Hence, the experimental functions (\ref{f++}) are simplified. Notice that we are giving the experimental functions of a qutrit for observables that are factorized in terms of two qubits. In \cite{bmmsm}, only factorizable density matrices are studied, but with the Bloch matrix method, an arbitrary density matrix is allowed.

\paragraph{Unitary  transformations.} It is worthy to stop here to see the form of (special) unitary transformations for the `nested' qutrit.

In \cite{zvsb}, a decomposition of the group $\rSU(4)$ (special unitary transformations of the two qubit system) is given in terms of a structure of{\it orthogonal symmetric Lie algebra} of the Lie algebra of $\rSU(4)$, $\mathfrak{su}(4)$. We  give here the definition \cite{hel}:

\begin{definition}{ \sl Orthogonal symmetric Lie algebra.} Let $\mathfrak{l}$ be a Lie algebra over $\R$ and $s$ an involutive automophism of $\mathfrak{l}$ (that is, $s^2=\rid$ and $s\neq \rid$). Let $\mathfrak u$ be the set of fixed points of $s$ and suppose that $\mathfrak{u}$ is  a compactly embedded subalgebra of $\mathfrak{l}$. Then, the pair $(\mathfrak{l}, s)$ is an orthogonal symmetric Lie algebra.

\hfill$\square$
\end{definition}
The Lie subalgebra $\mathfrak{u}$ is the eigenvector set of eigenvalue 1 of $s$. The eigenvector set of eigenvalue -1 is denoted by $\mathfrak{p}$ and it is, in general, just a vector space. One has
$$[\mathfrak{u}, \mathfrak{u} ]\subset \mathfrak{u},\quad [\mathfrak{u}, \mathfrak{p} ]\subset \mathfrak{p}, \quad [\mathfrak{p}, \mathfrak{p} ]\subset \mathfrak{u}.$$
The Lie algebra $\mathfrak{su}(4)=\{\hbox{ antihermitian, traceless, $4\times 4 $ matrices }\}$  has different structures of orthogonal symmetric Lie algebra. The structure that is of interest here is the one whose involutive automorphism is $s(X)=\bar X$. Then
\begin{align}
&\mathfrak{u}=\mathfrak{so}(4)=\{ \hbox{ real, antisymmetric $4\times 4$ matrices }\},\nonumber\\
&\mathfrak{p}=\{\hbox{ symmetric, purely imaginary $4\times 4$ matrices }\}.\label{Liealgebradecomposition}
\end{align}
A maximal abelian subspace of $\mathfrak{p}$ is the set of $4\times 4$ diagonal matrices.
\hyphenation{exhi-bi-ted}
\hyphenation{ge-ne-ral}
A basis of $\mathfrak{su}(4)$ is given in terms of  the Dirac matrices (\ref{diracmatrices})
\beq\left\{\,\frac \ri 2 \sigma_i\otimes \sigma_0,\quad \frac \ri 2\sigma_0\otimes\sigma_i,\quad \frac \ri 2\sigma_i\otimes\sigma_j,\quad i,j=1,2,3\,\right\}.\label{Diracbasis}\eeq These correspond to the  standard basis of $\C^2\otimes \C^2$ (\ref{standardbasis}), which is a basis of eigenvectors of $$\sigma_3\otimes \sigma_3=\begin{pmatrix}1&0&0&0\\0&-1&0&0\\0&0&-1&0\\0&0&0&1\end{pmatrix}.$$

Performing a change of basis from the standard one to the {\it Bell basis} (see for example \cite{zvsb})
\begin{align*}
&|b_1\rangle=\frac 1{\sqrt{2}}\left(|00\rangle+|11\rangle\right),\\
&|b_2\rangle=\frac \ri{\sqrt{2}}\left(|10\rangle+|01\rangle\right),\\
&|b_3\rangle= \frac 1{\sqrt{2}}\left(|10\rangle-|01\rangle\right),\\
&|b_4\rangle=\frac \ri{\sqrt{2}}\left(|00\rangle-|11\rangle\right),
\end{align*}
one can deduce directly the above decomposition. Let
$$B= \frac 1{\sqrt{2}}\begin{pmatrix}1&0&0&1\\
0&\ri&\ri&0\\0&1&-1&0\\\ri&0&0&-\ri\end{pmatrix}$$ and let
$$\tau_{\mu\nu}=B\sigma_{\mu\nu}B^+,$$
 Then, by an elementary calculation, one obtains
$$\mathfrak{u}=\rspan\left\{\tau_{01},\tau_{02},\tau_{03},\tau_{10},\tau_{20},\tau_{30}\right\},$$ since they are all real antisymmetric matrices. The rest
$$\mathfrak{p}=\rspan\left\{\tau_{11},\tau_{12},\tau_{13},\tau_{21},\tau_{22},\tau_{23},\tau_{31},\tau_{32},\tau_{33} \right\},$$ are symmetric, purely imaginary matrices. An abelian subset of $\mathfrak{p}$ is the subset of diagonal matrices
$$\mathfrak{a}=\rspan\{\tau_{11},\tau_{22},\tau_{33}\}.$$ This coincides with the decomposition given in \cite{zvsb} in the standard basis, where the commutation relations among the generators are explicitly exhibited.

On the other hand, it is well known the isomorphism $$\mathfrak{so}(4)\cong\mathfrak{so}(3)\oplus\mathfrak{so}(3)\cong\mathfrak{su}(2)\oplus \mathfrak{su}(2).$$
There is a theorem (see \cite{hel}, Chapter V, Theorem 6.7) that gives a decomposition of a group whose Lie algebra has the structure of an orthogonal symmetric Lie algebra.  This theorem, applied to our case, asserts  that $\rSU(4)$ admits a decomposition
\beq \rSU(4)\simeq \left(\rSU(2)\times \rSU(2)\right)\exp{\mathfrak{a}}\left(\rSU(2)\times \rSU(2)\right).\label{unitary}\eeq The factor $\exp{\mathfrak{a}}$ is a 3-torus.
A unitary transformation can then be cast in the form
$$U=k_1U(\alpha,\beta,\gamma)k_2,\qquad k_1, k_2\in \rSU(2)\times \rSU(2)\qquad \hbox{ with }$$
$$U(\alpha,\beta,\gamma)=\re^{(\alpha\tau_{11}+\beta\tau_{22}+\gamma\tau_{33})}=\re^{\alpha\tau_{11}}e^{\beta\tau_{22}}e^{\gamma\tau_{33}}=
U_1(\alpha)U_2(\beta)U_3(\gamma).$$
The transformations $k_1, k_2$  are interpreted as local transformations, that is, transformations involving, separately, the two subsystems of one qubit. The transformations $U(\alpha,\beta,\gamma)$ are instead  genuine non local transformations. In \cite{ga}, Appendix C, the general form of such $U(\alpha,\beta,\gamma)$ is given in terms of the Bloch matrix. For example,
$U(\alpha)$ will transform
$$[r_{\mu\nu}]=\begin{pmatrix}1&\vec{r}^t\\
\vec{s}&R\end{pmatrix}$$ into $$\vec{r'}=\begin{pmatrix}r_1\\r_2\cos\alpha +R_{13}\sin\alpha\\r_3\cos\alpha-R_{12}\sin\alpha\end{pmatrix},\qquad
\vec{s'}=\begin{pmatrix}s_1\\s_2\cos\alpha+ R_{31}\sin\alpha\\s_3\cos\alpha-R_{21}\sin\alpha\end{pmatrix},$$ and
$$R'=\begin{pmatrix}
R_{11}&r_3\sin\alpha+R_{12}\cos\alpha&-r_2\sin\alpha+R_{13}\cos\alpha\\
s_3\sin\alpha+R_{21}\cos\alpha&R_{22}&R_{23}
\\-s_2\sin\alpha+R_{31}\cos\alpha&R_{32}&R_{33}
\end{pmatrix}.$$
Such transformations preserve the qutrit condition (\ref{qutritcondition}), and so does a general transformation $U(\alpha,\beta, \gamma)$. Apparently, these transformations have three parameters, $\alpha, \beta$ and $\gamma$. Nevertheless, one can check with \cite{ga} that, if one assumes the qutrit condition, only  differences among these parameters occur. So we have a two dimensional abelian group $A$ of non local transformations. For completeness, we write here the transformations. We denote $\theta_1=\beta-\alpha$ and $\theta_2=\gamma-\alpha$. Then:
\begin{align*}
&r'_1=r_1\cos(\theta_1-\theta_2)+R_{23}\sin(\theta_1-\theta_2),\qquad &&r'_2=r_2\cos\theta_1-R_{13}\sin\theta_2,\\ &r'_3=r_3\cos\theta_1+R_{12}\sin\theta_1,\qquad &&R'_{12}=R_{12}\cos\theta_1+r_3\sin\theta_1,\\
&R'_{23}=R_{23}\cos(\theta_1-\theta_2)-r_1\sin(\theta_1-\theta_2),\qquad&&R'_{13}=R_{13}\cos\theta_2+r_2\sin\theta_2,
\\
&R'_{11}=R_{11},\qquad\qquad R'_{22}=R_{22},\qquad&& R'_{33}=R_{33}.
\end{align*}
The local transformations are reduced to $k_1,k_2\in\rSU(2)=\mathrm{diag}\left(\rSU(2)\times \rSU(2)\right)$,
so from (\ref{unitary}) we are left with a decomposition of unitary transformations of the qutrit as
$$ \rSU(2)\times  A   \times\rSU(2).$$
This is a subgroup of $\rSU(4)$ with 8 parameters. Since the unitary transformations for the qutrit are $\rSU(3)$, the formula above is a decomposition of $\rSU(3)$. The group $A$ is a 2-torus, isomorphic to the subgroup of $\exp(\mathfrak{a}) $ with $\alpha=0$.

At the infinitesimal level, the two (commuting) vectors fields, generators of the action of $A$ on the set of Bloch matrices, satisfying the qutrit condition, are
\begin{align*}
&\frac\partial{\partial \theta_1}=
R_{23}\frac\partial{\partial r_1}+
R_{12}\frac\partial{\partial r_3}+
r_3\frac\partial{\partial R_{12}}-r_1\frac\partial{\partial R_{23}},\\
&\frac\partial{\partial \theta_2}=-R_{23}\frac\partial{\partial r_1}
-R_{13}\frac\partial{\partial r_2}+
r_1\frac\partial{\partial R_{23}}+r_2\frac\partial{\partial R_{13}},
\end{align*}

\hfill$\blacksquare$

\section{ The fuzzy set picture}\label{fuzzy-sec}

Fuzzy sets were introduced by Zadeh \cite{zad} in a different context and soon they were used in many practical applications.

\begin{definition} \label{fuzzy-def} A fuzzy set  is a pair  $\cA=(\cU, \mu_\cA)$ where $\cU$ is a set called the universe of discourse and $\mu_\cA$ is a function
\begin{equation} \mu_\cA: \cU\rightarrow [0,1], \label{fuzzyset}\end{equation}
called the {\it membership function}. The number $\mu_\cA(x)$ indicates the {\it degree of membership} with which the element $x$ belongs to $\cA$, 0 meaning no membership and 1 full membership. We will use in indistinctly the notation $\cA$ or $\mu_\cA$ to refer to a fuzzy set.

\hfill $\square$

\end{definition}

\begin{definition} Let $\cA$ and $\cB$ be two fuzzy sets in the same universe of discourse $\cU$.

\begin{enumerate}
\item $\cA=\cB$ if an only if $\mu_\cA(x)=\mu_\cB(x)$ $\;\forall x\in \cU$.
\item $\cA\subseteq\cB$ if an only if $\mu_\cA(x)\leq\mu_\cB(x)$ $\;\forall x\in \cU$.
\item The {\it bold union} $\cA\sqcup\cB$  is defined as
$$\mu_{\cA\sqcup\cB}(x)=\mathrm{min}\{\mu_\cA(x)+\mu_\cB(x), 1\}.$$
\item The {\it bold intersection} $\cA\sqcap \cB$  is defined as
$$\mu_{\cA\sqcap\cB}(x)=\mathrm{max}\{\mu_\cA(x)+\mu_\cB(x)-1, 0\}.$$
\item The {\it  complementation} of $\cA$ is
$$\mu_{\cA'}(x)=1- \mu_\cA(x) \qquad \forall x\in \cU.$$
\item One defines the empty set $\emptyset$ by $\mu_\emptyset(x)=0$, and its complement is $\cU$ itself with $\mu_\cU(x)=1$
\item $\cA$ and $\cB$ are {\it weakly disjoint} if ${\cA\sqcap\cB}=\emptyset$.

\end{enumerate}

\hfill $\square$

\end{definition}
\hyphenation{ex-pe-ri-men-tal}

The bold union and intersection \cite{gi,zaw} are not the union and intersection originally introduced by Zadeh, that is,

\begin{align}
\mu_{\cA\cup\cB}(x)=\mathrm{max}\{\mu_\cA(x),\mu_\cB(x)\},\nonumber\\
\mu_{\cA\cap\cB}(x)=\mathrm{min}\{\mu_\cA(x),\mu_\cB(x)\}.\label{uizadeh}\end{align}
The union and intersection (\ref{uizadeh})  do not satisfy the law of excluded middle and the law of contradiction,
$$\cA\sqcap{\cA'}= \emptyset,\qquad\cA\sqcup {\cA'}= \cU,$$ but they are distributive (see Equation (\ref{distributivity})); then they are not well suited to describe a quantum system.

Instead, the bold union and intersection satisfy the law of excluded middle, the law of contradiction and are not distributive.

On orthogonal sets,  $\cA\sqcap \cB =\emptyset$,  one can see that the bold union is just

$$\mu_{\cA\sqcup\cB}(x)=\mu_\cA(x)+\mu_\cB(x).$$
Our aim is to interpret them in the context of the  Birkhoff-von Neumann (standard) logic.

\medskip

\hyphenation{pro-per-ties}

In the fuzzy approach to Quantum Mechanics by Pykacz \cite{py1,py2}, the experimental functions of Theorem \ref{mac-theorem} are interpreted as membership functions of fuzzy sets in the universe of discourse formed by all the possible states of the quantum system. We quote here the theorem of \cite{py1,py2} that allows such interpretation.

\begin{theorem}\label{pykacz-theo}
Any logic $L$ with an ordering  set of probability measures $\cS$ can be mapped isomorphically to a logic formed by a family of fuzzy subsets of $\cS$, $\cL^f(\cS)$, that satisfies the following properties:
\begin{enumerate}

\item $\emptyset$ belongs to the family $\cL^f(\cS)$.
\item If $\cA$ belongs to $\cL^f(\cS)$, so does its complement $\cA'$.
\item Given a countable family  $\{\cA_i \}_{i\in I}$  of fuzzy subsets inside $\cL^f(\cS)$ which is pairwise weakly disjoint, $\cA_i\sqcap\cA_j=\emptyset$ for $i\neq j$, then $\sqcup _{i\in I}\cA_i$ belongs to $\cL^f(\cS)$.
\item If $\cA\sqcap \cA=\emptyset$ then $\cA=\emptyset$.

\end{enumerate}

Conversely, any family of fuzzy subsets of a universe $\cU$, satisfying properties 1 - 4 above, is a logic where the  partial order is the inclusion of fuzzy sets and the orthocomplementation is the complementation of fuzzy sets. The notion of orthogonality coincides with that of weak disjointness.

The set of ordering measures is given by elements of the universe $\cU$ as
\begin{equation}m_\rho(\cA)=\mu_\cA(\rho)\qquad \forall \rho \in \cU\, .\label{probability measure}\end{equation}

\hfill $\square$

\end{theorem}

The universe of discourse $\cU$ is the set of states $\cS$.  As stated before, given two membership functions $\mu_\cA(\rho)$ and $\mu_\cB(\rho)$, the bold intersection being the empty set, implies that the bold union is
$$\mu_\cA\sqcup\mu_\cB =\mu_\cA+\mu_\cB\leq 1,$$
and then, the function $m_\rho:\cL^f\rightarrow[0,1]$ above satisfies the conditions to be a probability measure.
The same is true for an infinite countable succession of $\cA_i$; this is guaranteed by property 3 in Theorem \ref{pykacz-theo}.

\medskip

In Quantum Mechanics,
each state is represented by a density matrix, therefore the universe of discourse is $$\cS=\left \{\, \rho\in M_{n\times n}(\C)\; |\; \rho=\rho^\dag,\, \rho\hbox{ is non negative and } \tr\rho=1\,\right\}.$$ By stating this we have already introduced the Hilbert space. The membership functions (\ref{experimentalfunction1}) are determined by an observable $A$ and a Borel set $E\subset \R$
$$f^A_E(\rho)=p(A,\rho, E)=\tr(\rho P^A_E),$$
where $p(A,\rho, E)$ is the probability map of Mackey (\ref{probability map}) and $P_E^A$ is the projector on the subspace of $\cH$ that is the sum of the eigenspaces of $A$ in $\cH$ with eigenvalue in the Borel set $E$.  So the membership function $\mu_\cA(\rho)=f_A^E(\rho)$  contains information about the observable $A$ and the Borel subset $E$.

\subsection{The fuzzy bit}

For the qubit, the membership (experimental) functions (\ref{experimentalfunction}) are labelled by a unit vector $\hat{a}$
$$\mu_\cA=f_{\hat{a}}(\vec{\rho})=\frac 12 +\hat{a}\cdot \vec{\rho}=\frac 12 +|\vec{\rho}|\cos\theta,$$
where $$\rho=\frac12\rid+\rho_i\sigma_i=\frac12\rid+\vec{\rho}\cdot\vec{\sigma},$$ plus the constant functions 0 and 1. 

They
are functions $f:\cU\rightarrow[0,1]$ that, according to Theorem \ref{pykacz-theo}, can be interpreted as membership functions of fuzzy sets. These functions are orthogonal if the bold intersection of fuzzy sets is the empty set, so
$$\mu_\cA\sqcap\mu_\cB=\mathrm{max}\{\hat{a}\cdot \vec{\rho}+ \hat{b}\cdot \vec{\rho}, 0\}=0\quad \Leftrightarrow \quad\hat{a}\cdot \vec{\rho}+ \hat{b}\cdot \vec{\rho}<0.$$ Since this relation has to be satisfied for all $\rho\in \cS$ we recover the result in  (\ref{orthogonalityqubit}), that is, $\hat{a}=-\hat{b}$.

If the fuzzy sets are orthogonal, then its bold union is

$$\mu_\cA\sqcup \mu_\cB=\mathrm{min}\{\mu_\cA+\mu_{\cB}, 1\}=1,$$ that is, the union is the universe of discourse $S$, which makes sense, since it is the probability that the eigenvalue of $A$ is +1 or -1, which exhausts all the possibilities.

We now give some examples that help to clarify the role of the fuzzy sets for the qubit.

\hyphenation{or-tho-go-na-li-ty}

\begin{example}{\sl The gate NOT.} We consider a qubit. Let $\{\,|0\rangle,|1\rangle\,\}$ be the ordered basis of $\cH\simeq\C^2$ of eigenvectors of $\sigma_3$,
$$\sigma_3 =\begin{pmatrix}1&0\\0&-1\end{pmatrix} .$$ Let $|\Psi\rangle=v_0|0\rangle+v_1|1\rangle\in \cH$ a vector representing a pure state, so $|v_0|^2+|v_1|^2=1$. There is a transformation that flips the two basis states (the NOT gate). It is represented in this basis by the unitary matrix
$$U=\begin{pmatrix}0&1\\1&0\end{pmatrix}=\sigma_1,\qquad U|\Psi\rangle=v_1|0\rangle+v_0|1\rangle.$$
Notice that $\det U=-1$.
Its action on a density matrix is
$$\rho=\begin{pmatrix}\rho_{00}&\rho_{01}\\{\bar\rho_{01}}&\rho_{11}\end{pmatrix}\, \longrightarrow\, U\rho U^\dagger=\begin{pmatrix}\rho_{11}&{\bar\rho}_{01}\\\rho_{01}&\rho_{00}\end{pmatrix}.$$
In terms of the sigma matrices we have that
$$U\sigma_1U^\dagger=\sigma_1,\qquad U\sigma_2U^\dagger=-\sigma_2,\qquad U\sigma_3U^\dagger=-\sigma_3,$$
so the flip transformation takes $\vec{\rho}=(\rho_1,\rho_2,\rho_3)$ to $\vec{\rho}{\phantom{'}}'=(\rho_1,-\rho_2,-\rho_3)$. In the Bloch sphere, this is a rotation of $180^0$ around the $x_1$ axis.

The probability of finding the system in a state `+' with respect to an arbitrary axis $\hat{a}$ is
$$f_{\hat a}^+(\rho)=\frac 12 + \hat{a}_1\rho_1+\hat{a}_2\rho_2+\hat{a}_3\rho_3.$$ Effecting the NOT transformation  changes the probability to
\beq f_{\hat a}^+(\rho')=\frac 12 +\hat{a}_1\rho_1-\hat{a}_2\rho_2-\hat{a}_3\rho_3.\label{not}\eeq
Notice that applying the NOT gate to the fuzzy bit is not the same than computing its complement ${f_{\hat a}^+(\rho)}'= 1-f_{\hat a}^+(\rho)$ as a fuzzy set. This has been cause of some confusion in the literature. However, in the particular case  that  $\hat a$ is the $x_3$ axis,  $\hat a=(0,0,1)$, then one has that
$$f_{\hat a}^+(\rho')=\frac 12-\rho_3,$$
which is equal to its complement as a fuzzy set
$$1-f_{\hat a}^+(\rho)=\frac 12-\rho_3.$$ This probability coincides with $f_{\hat{a}}^-(\rho)=f_{-\hat{a}}^+(\rho)$.

\hfill$\square$
\end{example}

\begin{example}{\sl The gate $\sqrt{\hbox{NOT}}$.} In the same setting than the example above, one defines the square root of the gate NOT, $\sqrt{\hbox{NOT}}$, as a unitary transformation such that
\beq\sqrt{\hbox{NOT}}\cdot\sqrt{\hbox{NOT}}=\hbox{NOT}.\label{sqrtnot}\eeq This transformation is not unique, but it is usual to choose the following one:
$$U=\frac 12 \begin{pmatrix}1+\ri&1-\ri\\1-\ri&1+\ri\end{pmatrix}.$$ (Notice that $\det U=\ri$.) The  sigma matrices behave, under this transformation, as
$$U\sigma_1U^\dagger=\sigma_1,\qquad U\sigma_2U^\dagger=\sigma_3,\qquad U\sigma_3U^\dagger=-\sigma_2,$$ so $\vec{\rho}=(\rho_1,\rho_2,\rho_3)$ goes to $\vec{\rho}{\phantom{'}}'=(\rho_1,\rho_3,-\rho_2)$. In the Bloch sphere, this is a rotation of $90^0$ around the $x_1$ axis. It is clear that (\ref{sqrtnot}) is satisfied. The membership functions then change as

$$f_{\hat a}^+(\rho')=\frac 12 +\hat{a}_1\rho_1+\hat{a}_2\rho_3-\hat{a}_3\rho_2.$$

\bigskip

It is known \cite{cgg, kkk} that the negation in fuzzy logic (orthocomplementation of fuzzy sets) doesn't have a square root. Particularly,
Lemma 17.1.12 in \cite{cgg} reads: ``There is no continuous function $f : [0, 1] \rightarrow[0, 1]$ such
that for any $x \in [0, 1]$, $f(f(x)) = 1-x$''. The authors then provide an example of square root in which the function $f$ is not continuous.

In our approach, since the NOT gate sending the fuzzy set  $ f_{\hat a}^+(\rho)$ to ${f_{\hat a}^+(\rho)}'=f_{\hat a}^+(\rho')$ does not coincide with the negation, ${f_{\hat a}^+(\rho)}'\neq 1-f_{\hat a}^+(\rho)$ in general, a continuous square root of the gate NOT is allowed.

\hfill$\square$
\end{example}
\subsection{The entangled fuzzy bits}
For the entangled fuzzy bits, the membership functions of $\mathcal{E}=\cA\times\cB$ are given in terms of the two unit vectors $\hat {a}$ and $\hat{b}$ (\ref{f++}):
$$ \mu_\mathcal{E}(\vec{s},\vec{r}, R)=f^{+,+}_{\hat{a},\hat{b}}(\vec{s},\vec{r}, R)=\frac 14 \left( 1+\vec{s}\cdot \hat{a}+\vec{r}\cdot \hat{b}+\hat{a}^{\, t}R\hat{b}\right).$$ In the two examples (\ref{inequality00}) and (\ref{inequality0}) one can check directly that the orthogonality of the membership functions implies that the bold union is again a membership function:
\begin{align*}&f^{+,+}_{\hat{a},\hat{b}}(\vec{s},\vec{r}, R)+f^{+,-}_{\hat{a},\hat{b}}(\vec{s},\vec{r}, R)=f^{+,+-}_{\hat{a},\hat{b}}(\vec{s},\vec{r}, R)\leq 1,\\
&f^{+,+}_{\hat{a},\hat{b}}(\vec{s},\vec{r}, R)+f^{-,-}_{\hat{a},\hat{b}}(\vec{s},\vec{r}, R)\leq 1,
\end{align*}
that is, the sum is again a probability.

\begin{example} {\sl The CNOT gate.} The Controlled NOT gate has an input of two qubits. In the ordered basis of $\cH\simeq \C^2\otimes\C^2$ used above, the CNOT transformation acts as
$$\begin{CD}\{|00\rangle,|01\rangle,|10\rangle,|11\rangle\}\}@>\hbox{CNOT}>>\{|00\rangle,|01\rangle,|11\rangle,|10\rangle\}\end{CD}$$It is represented in this basis by the unitary matrix
$$U=\begin{pmatrix}1&0&0&0\\0&1&0&0\\0&0&0&1\\0&0&1&0\end{pmatrix}=\begin{pmatrix}\rid&0\\0&\sigma_1\end{pmatrix} .$$
Notice that $\det U=-1$.
We want to see how the membership functions change with this transformation, so we will compute first the change in the density matrix. Let us remind the notation (\ref{decomposition}):
$$\rho=\frac 14r_{\mu\nu}\sigma_\mu\otimes \sigma_\nu, \qquad \hbox{with}\qquad[r_{\mu\nu}]=
\begin{pmatrix}
1&r_{01}&r_{02}&r_{03}\\
r_{10}&r_{11}&r_{12}&r_{13}\\
r_{20}&r_{21}&r_{22}&r_{23}\\
r_{30}&r_{31}&r_{32}&r_{33}
\end{pmatrix}=\begin{pmatrix}1&{\vec{r}\phantom{'}}^{t}\\\vec{s}&R\end{pmatrix}.$$
 For the transformed matrix
$$\rho'=U\rho U^\dagger=\frac 14r'_{\mu\nu}\sigma_\mu\otimes \sigma_\nu,\qquad[r'_{\mu\nu}]=\begin{pmatrix}1&{\vec{r}\phantom{'}'}^{t}\\\vec{s}\phantom{'}'&R'\end{pmatrix},$$
we obtain \cite{MATHEMATICA}:
\begin{align*}
&{\vec{r}\phantom{'}'}^{t}=(r_{01},r_{32},r_{33}),&&{\vec{s}\phantom{'}'}^{t}=(r_{11},r_{21},r_{30}),\\&R'=
\begin{pmatrix}
r_{10}&r_{23}&-r_{22}\\
r_{20}&-r_{13}&r_{12}\\
r_{31}&r_{02}&r_{03}
\end{pmatrix}, &&r'_{00}=r_{00}=1.
\end{align*}
Let us consider $A=B=\sigma_3$, so $\hat{a}=\hat{b}=(0,0,1)$. Suppressing $\hat{a}$ and $\hat{b}$ from the notation in (\ref{f++}), we get
\begin{align*}
&f^{++}=\frac 14 (1+r_{30}+r_{03}+r_{33}),&& {f'}^{++}=\frac 14 (1+r_{30}+r_{33}+r_{03}),\\
&f^{+-}=\frac 14 (1+r_{30}-r_{03}-r_{33}),&& {f'}^{+-}=\frac 14 (1+r_{30}-r_{33}-r_{03}),\\
&f^{-+}=\frac 14 (1-r_{30}+r_{03}-r_{33}),&& {{f'}^{-+}}=\frac 14 (1-r_{30}+r_{33}-r_{03}),\\
&f^{--}=\frac 14 (1-r_{30}-r_{03}+r_{33}),&& {f'}^{--}=\frac 14 (1-r_{30}-r_{33}+r_{03}).
\end{align*}
As expected, since the first qubit is a control qubit, we have
\begin{align*}
&f^{++}={f'}^{++}, &&f^{+-}={f'}^{+-}, && f^{-+}={f'}^{--}, &&f^{--}={f'}^{-+}.
\end{align*}

\hfill$\square$
\end{example}

\section{Conclusions}\label{conclusion-sec}

In this paper the meaning of the relationship between Quantum Mechanics, logics and fuzzy sets has been clarified. It results that a particular family of fuzzy sets, with universe of discourse the density matrices in a Hilbert space, can actually represent faithfully the Birkhoff-von Neumann logic of projectors of a Hilbert space. This choice is the same than the `{\it ad hoc}' choice of Mackey in his Axiom VII. The membership functions of the fuzzy sets (or, equivalently, the experimental functions of   M\c{a}czy\'{n}ski) have been computed for three different systems: the qubit,  two entangled qubits and a qutrit `nested' inside the entangled qubits. A decomposition of SU(3), the special unitary transformations of the qutrit, inherited from the orthogonal symmetric Lie algebra structure of $\fsu(4)$, has been found. With the aid of the membership functions, we have proven that a continuous square root of NOT exists for the qubit.

\section*{Acknowledgements}
This work is supported by the Spanish Grants FIS2017-
84440-C2-1-P funded by MCIN/AEI/10.13039/501100011033 ‘ERDF A way of making
Europe’, PID2020-116567GB-C21 funded by MCIN/AEI/10.13039/501100011033, the
Project PROMETEO/2020/079 (Generalitat Valenciana) and the project CaLIGOLA, Project ID: 101086123 (HORIZON).

This article is based upon work from COST Action CaLISTA CA21109 supported by COST (European Cooperation in Science and Technology). www.cost.eu.

M. Aldana wants to thank the Departament de F\'{\i}sica Te\`{o}rica, Universitat de Val\`{e}ncia, for its kind hospitality.

\end{document}